\documentclass[%
 aip,
 pof,%
 amsmath,amssymb,
preprint,%
]{revtex4-1}

\usepackage{graphicx}% Include figure files
\usepackage{dcolumn}% Align table columns on decimal point
\usepackage{bm}% bold math
\begin{document}

%\preprint{AIP/123-QED}

\title[Biased swimming cells do not disperse in pipes as tracers]{Biased swimming cells do not disperse in pipes as tracers: a population model based on microscale behaviour}% Force line breaks with \\
%\thanks{Footnote to title of article.}

\author{R. N. Bearon}
\email[Corresponding author ]{rbearon@liv.ac.uk}
\affiliation{
Department of Mathematical Sciences, University of Liverpool, Peach Street, Liverpool, L69 7ZL, UK
}%
\author{M. A. Bees }%
\author{O. A. Croze}
\affiliation{%
School of Mathematics and Statistics, University of Glasgow, Glasgow, Scotland, G12 8QW, UK
}%

\date{\today}% It is always \today, today,
             %  but any date may be explicitly specified

\begin{abstract}
There is much current interest in modelling suspensions of algae and other micro-organisms for biotechnological exploitation, and many bioreactors are of tubular design.
Using generalized Taylor dispersion theory, we develop a population-level swimming-advection-diffusion model for suspensions of  micro-organisms in a vertical pipe flow.  In particular, a combination of gravitational and viscous torques acting on individual cells can affect their swimming behaviour, which is termed gyrotaxis.  This typically leads to local cell drift and diffusion in a suspension of cells.
In a flow in a pipe, small amounts of radial drift across streamlines can have a major impact on the effective axial drift 
and diffusion of the cells.
We present a Galerkin method to calculate the local mean swimming velocity and diffusion tensor based on local shear for arbitrary flow rates. This method is validated with asymptotic results obtained in the limits of weak and strong shear.   We solve the resultant swimming-advection-diffusion equation using numerical methods for the case of imposed Poiseuille flow and investigate how the flow modifies the dispersion of active swimmers from that of passive scalars.  We establish that generalized Taylor dispersion theory predicts an enhancement of gyrotactic focussing in pipe flow with increasing shear strength, in contrast to earlier models.  We also show that biased swimming cells may behave very differently to passive tracers, drifting axially at up to twice the rate and diffusing much less.
\end{abstract}

% PACS, the Physics and Astronomy
                             % Classification Scheme.
%Use showkeys class option if keyword
                              %display desired
\maketitle

\section{Introduction}

Swimming micro-organisms, such as algae and bacteria, have their own agenda;
selective pressures lead cells to adopt strategies to optimize a combination of environmental 
conditions, such as illumination, nutrients or the exchange of genetic material.
This can significantly impact the behaviour of suspensions of swimming micro-organisms,
particularly in flows where biased motion across streamlines can lead to rapid transport.
For example, various algae are gravitactic, that is they swim upwards on average in still 
fluid which can be beneficial for reaching regions of optimal light. For some species 
this is due to being bottom-heavy - the centre of gravity for these cells is offset from the centre of buoyancy, and the combination of the effects of gravity with the buoyancy force gives rise to a gravitational torque which serves to reorient the cell allowing it to swim upwards - whereas in others sedimentary torques lead to similar behaviour \citep{Roberts:2006}.  
However, in shear flow the cells may be reoriented from the vertical due to 
viscous torques \citep{Pedley:1992a}.  For a vertical pipe containing downwelling fluid,
gravitactic cells can accumulate near the centre \citep{Kessler:1985a}, a 
phenomenon known as gyrotactic focussing.  As recently predicted theoretically by \citet{Bees:2010}, such a modification of the spatial distribution 
of algae in tubes alters significantly the effective axial dispersion of the 
cells.

There is much current interest in employing micro-organisms for biotechnological
purposes, from the production of biofuels\cite{Melis:2001,Chisti:2007}, such as 
hydrogen, biomass or lipids, to high-value products, such as $\beta$-carotene.
Cells are grown either extensively on low value land or intensively to optimize
growth.  Intensive culture systems typically consist of arrays of tubes (vertical, horizontal
or helical) and aim to maximize light and nutrient uptake.  Bioreactors may
be pumped or bubbled, in turbulent or laminar regimes.  However, energy input 
may be energy wasted; efficient bioreactor designs might aim to make use of the
swimming motion of the cells themselves, or accommodate the fact that swimming micro-organisms
(where drift across streamlines is more important than axial motion) and 
nutrients are likely to drift and diffuse at different rates along the tubes.

In a still fluid, the swimming behaviour of individual gyrotactic phytoplankton has been usefully 
described as a biased random walk: the cell orientation is assumed to be a random 
variable that undergoes diffusion with drift \citep{Hill:1997}. At the population-level 
the dynamics can be modelled with a swimming-diffusion equation for the cell 
concentration, where the cells swim in a preferred direction
at a mean velocity and diffuse with an anisotropic diffusion tensor that represents the 
random component of swimming \citep{Bearon:2008}. Extending such population-level models 
to incorporate the effects of ambient flow is non-trivial. Although the orientation 
distribution and resultant mean swimming velocity of such cells in unbounded homogeneous 
shear flow has previously been computed \cite{Bees:1998a,Almog:1998}, the resultant 
diffusion tensor is more complicated.  For homogeneous shear flow, subject to certain 
constraints on the form of the flow, \citet{Hill:2002} and \citet{Manela:2003} calculated 
expressions for the diffusion tensor using the theory of generalized Taylor dispersion (GTD).
Because of its account of shear-induced correlations in cell position, GTD is a more rational account over earlier approaches based on an orientation only description
using a Fokker-Planck equation and diffusion tensor estimate (FP)\cite{Pedley:1990}.

\citet{Bearon:2011} compared two-dimensional individual-based simulations of 
swimming micro-organisms with swimming-advection-diffusion models for the
whole population in situations where the flow is not homogeneous, 
that is in flows in which the cells can experience a range of shear environments. 
Using GTD theory to calculate local expressions for the mean swimming direction and 
diffusion coefficients, the results of the individual and population models were
generally in good agreement and were able to successfully predict the phenomena 
of gyrotactic focussing. However, this work was restricted to two-dimensions;
both the swimming motions and velocity field were confined to a vertical plane.

Here, we consider axisymmetric pipe flow, which locally can be described by planar 
shear, and consider swimming motions which are allowed to be fully three-dimensional. 
First, we develop a population-level swimming-advection-diffusion model where the 
mean swimming velocity and diffusion tensor are based on the local shear.  
Next, a Galerkin method is presented for calculating the mean swimming velocity 
and diffusion tensor based on the local shear, and asymptotic results are obtained 
in the limits of weak and strong shear.  
The resultant swimming-advection-diffusion equation is then solved numerically 
for the case of imposed Poiseuille flow.  We contrast the GTD results with the
FP approach.
Finally, we investigate how the flow modifies qualitatively and quantitatively 
the dispersion of active swimmers from that of a passive scalar.

This paper represents an important link study that will facilitate the comparison
of the exact long-time theoretical results of \citet{Bees:2010}
and the forthcoming experimental results by the authors on the transient dynamics.

\section{Mathematical Model}

\subsection{Vertical pipe flow}\label{sec:vert}
Consider axisymmetric fluid flow with velocity $\mathbf{u}$ through a vertical tube of circular cross-section, radius $a$,  with axis parallel to the $z$-axis pointing in the downwards direction, such that
\begin{eqnarray}
\label{eq:pipe_flow}
\mathbf{u}=u(r)\mathbf{e}_z=U(1+\chi(r/a))\mathbf{e}_z.
\end{eqnarray}
Here, $U$ is the mean flow speed, $U\chi$ is the variation of the flow speed relative to the mean, $r$ is the radial distance from the centre of the tube and ($\mathbf{e}_r,\mathbf{e}_\psi,\mathbf{e}_z$) are right-handed orthonormal unit vectors that define the cylindrical co-ordinates.  For flow subject to a uniform pressure gradient and no-slip boundary conditions on the walls, we have simple Poiseuille flow, $\chi(r)=1-2r^2$. In the fully coupled problem, where the negative buoyancy of the cells modifies the flow, $\chi(r)$ must be determined, as in \citet{Bees:2010}.

A population-level model for gyrotactic micro-organisms in {\em homogeneous} shear flow has previously been derived based on generalized Taylor dispersion theory\cite{Hill:2002,Manela:2003} (GTD). Specifically, for particular types of flow and on timescales long compared to $1/d_r$,  where $d_r$ is the rotational diffusivity due to the intrinsic randomness in cell swimming, the cell concentration $n(\mathbf{x},t)$ was shown to satisfy a swimming-advection-diffusion equation of the form
\begin{eqnarray}
\label{eq:ad_diff_in_flow}
\frac{\partial n}{\partial t}+\nabla_\mathbf{x}.\left[\left(\mathbf{u}+V_s\mathbf{q}\right)n-\frac{V_s^2}{d_r}\mathbf{D}.\nabla_\mathbf{x}n\right]=0,
\end{eqnarray}
where $V_s$ is the constant cell swimming speed, and $\mathbf{q}$ and $\mathbf{D}$ are the non-dimensional mean cell swimming direction and diffusion tensor, respectively. Explicit expressions for $\mathbf{q}$ and $\mathbf{D}$ as a function of the local shear strength will be given in section \ref{sec:GTD}.
Furthermore, \citet{Bearon:2011} show that this population-level approach is a good approximation 
for flow fields more general than homogeneous shear.  Therefore, we shall use (\ref{eq:ad_diff_in_flow}) to describe the 
cell concentration in a pipe flow with non-homogeneous shear. To solve the swimming-advection-diffusion 
equation numerically, it is convenient to non-dimensionalize lengths based on the pipe radius, $a$, 
and non-dimensionalize time on $a^2d_r/V_s^2$, a characteristic timescale for diffusion across the 
pipe.  This reveals two non-dimensional parameters in the problem: the P\'{e}clet number which 
is given by
\begin{eqnarray}
Pe&=&\frac{Ua d_r}{{V_s}^2},
\end{eqnarray}
and $\beta$, the ratio of  pipe radius to a typical correlation length-scale of the random walk in the absence of bias, defined as
\begin{eqnarray}
\label{eq:beta}
\beta=\frac{ ad_r }{V_s}.
\end{eqnarray}
An alternative interpretation of $\beta=aV_s/(V_s^2d_r^{-1})$ is as a `swimming P\'eclet number'.  Equation (\ref{eq:ad_diff_in_flow}) in non-dimensional form thus becomes
\begin{eqnarray}
\label{eq:nd_ad_diff_in_flow}
\frac{\partial n}{\partial t}+\nabla_\mathbf{x}.\left[(Pe[1+\chi(r)]\mathbf{e}_z+ \beta\mathbf{q})n-\mathbf{D}.\nabla_\mathbf{x} n\right]=0.
\end{eqnarray}

\subsection{Generalized Taylor dispersion}
\label{sec:GTD}

The shear in the pipe flow given by (\ref{eq:pipe_flow}) can be locally described as a simple shear flow. Specifically, consider a Taylor expansion of the flow field near some reference point $\mathbf{R}_0$ which is at radial position $r=R_0$
\begin{eqnarray}
\mathbf{u}(\mathbf{R}) \approx \mathbf{u}(\mathbf{R_0})+(\mathbf{R}-\mathbf{R}_0).\mathbf{e}_r \frac{U}{a}\chi'(R_0/a)\mathbf{e}_z.
\end{eqnarray}
We consider local co-ordinates relative to an origin located at $\mathbf{R}_0$ such that   $\mathbf{k}$  is pointing vertically upwards and ($\mathbf{i},\mathbf{j},\mathbf{k}$)  form a right-handed orthonormal set of unit vectors so that
\begin{eqnarray}
\label{eq:local_co_ord1}
\mathbf{i} =\mathbf{e}_r,\quad \mathbf{j}=-\mathbf{e}_\psi,\quad \mathbf{k}=-\mathbf{e}_z.
\end{eqnarray}
Defining the local position co-ordinate, $\mathbf{R}-\mathbf{R}_0=\xi \mathbf{i}+\eta\mathbf{j}+\zeta\mathbf{k}$, the flow field can then be written locally as simple shear, such that
\begin{eqnarray}
\mathbf{u}(\mathbf{R})=\mathbf{u}(\mathbf{R}_0)+ G\xi\mathbf{k},
\end{eqnarray}
where the shear strength $G$ is given by  $-\frac{U}{a}\chi'$. With this choice of co-ordinates, the velocity gradient tensor, $\mathbf{G}$, defined such that $\mathbf{u}(\mathbf{R})=\mathbf{u}(\mathbf{R}_0)+(\mathbf{R}-\mathbf{R}_0).\mathbf{G}$, has the simple form $G_{ij}=G \delta_{i1}\delta_{j3}$.

The mean swimming direction, $\mathbf{q}$, and non-dimensional diffusion tensor $\mathbf{D}$ can be written as integrals over cell orientation, $\mathbf{p}$, in the form \cite{Hill:2002,Manela:2003}
\begin{eqnarray}
 \label{eq:def_p}
\mathbf{q}&=&\int_\mathbf{p}\mathbf{p}  f(\mathbf{p} ) d\mathbf{p} ,\\
\label{eq:pos_def_diffusion}
\mathbf{D}&=&\int_\mathbf{p} [\mathbf{b}\mathbf{p}+\frac{2\sigma}{ f(\mathbf{p})}\mathbf{b}\mathbf{b}.\hat{\mathbf{G}}]^{sym}d\mathbf{p}.
\end{eqnarray}
Here $[]^{sym}$ denotes the symmetric part of the tensor, $\hat{\mathbf{G}}=\mathbf{i}\mathbf{k}$, and  $\sigma$ is a non-dimensional measure of the shear, defined as
\begin{eqnarray}
\label{eq:Pr_Pe_beta_relate}
\sigma=\frac{ G }{2d_r}=-\frac{Pe }{2 \beta^2} \chi'.
\end{eqnarray}
We note that $\sigma$ varies with $r$ because the shear varies across the radius of the tube. However, in the theory of GTD the shear is assumed locally homogeneous, and so we calculate local expressions for the mean swimming and diffusion based on the local value of $\sigma$.

The equilibrium orientation, $f(\mathbf{p})$, and vector $\mathbf{b}(\mathbf{p})$ satisfy\cite{Hill:2002,Manela:2003}
\begin{eqnarray}
\label{eq:f_eqn_with_flow}
\mathcal{L}f&=&0,\\
\label{eq:b_eqn_with_flow}
\mathcal{L}\mathbf{b}-2\sigma\mathbf{b}.\hat{\mathbf{G}}&=&f(\mathbf{p})(\mathbf{p}-\mathbf{q}),
\end{eqnarray}
subject to the integral constraints
\begin{eqnarray}
\label{eq:int_constraints}
\int_\mathbf{p} f d\mathbf{p}=1,\quad \label{eq:fnorm}
\int_\mathbf{p} \mathbf{b}d\mathbf{p}=0.\label{eq:bnorm}
\end{eqnarray}
Here, the linear operator $\mathcal{L}$ for a spherical swimming cell is defined by
\begin{eqnarray}
\label{eq:mathcalG}
\mathcal{L}f&=&
\nabla_\mathbf{p}.(
(\lambda(\mathbf{k}-(\mathbf{k}.\mathbf{p})\mathbf{p})-\sigma\mathbf{j} \wedge \mathbf{p}
)f-\nabla_\mathbf{p} f),
\end{eqnarray}
the gyrotactic bias in swimming direction is represented by the non-dimensional parameter 
\begin{eqnarray}
\label{eq:def_lambda}
\lambda=\frac{1}{2d_rB},
\end{eqnarray}
and $B = \mu\alpha_{\perp}/2h\rho g$ is the gyrotactic reorientation time scale, where $h$ is the distance between an average cell's centre-of-mass and centre-of-buoyancy, $\alpha_{\perp}$ is the dimensionless resistance coefficient for rotation about an axis
perpendicular to $\mathbf{p}$, $\mu$ and $\rho$ are the fluid viscosity and density respectively, and $g$ is the magnitude of the gravitational force.

To summarize, the non-dimensional mean swimming velocity and diffusion tensor are given as functions of two non-dimensional parameters: $\lambda$, which only depends on properties of the cell, and $\sigma$, which quantifies the strength of the shear.  See equations (\ref{eq:def_p}-\ref{eq:def_lambda}).
%  (equations \ref{eq:def_p}, \ref{eq:pos_def_diffusion}, \ref{eq:f_eqn_with_flow}-\ref{eq:mathcalG}).
Furthermore, for the pipe flow considered in the previous section, we can express $\sigma$ as a simple function of the non-dimensional parameters $Pe$, the global P\'eclet number, and $\beta$, the swimming P\'eclet number, and the shear profile $\chi'(r)$ (Eq. \ref{eq:Pr_Pe_beta_relate}).  The solution of the governing non-dimensional swimming-advection-diffusion equation (Eq. \ref{eq:nd_ad_diff_in_flow}) can therefore be determined by specifying the three non-dimensional parameters $\lambda, Pe$ and $\beta$ and the non-dimensional flow profile $\chi (r)$.

When explicit calculations are presented in this paper we have assumed that the flow is Poiseuille, $\chi(r)=1-2r^2$, and take $\lambda=2.2$ so as to compare with previous work \cite{Hill:2002,Bees:1998a} based on the algal species {\it C. augustae} (wrongly identified as {\it C. nivalis}\cite{Croze:2010}).  In \citet{Bearon:2011}, good agreement was found in planar pipe flow between individual based simulations and the population-level model with $\beta=10$ (where the reciprocal of $\beta$ was defined as $\epsilon=0.1$ therein). Motivated by the pipe dimensions in experiments currently in progress, we also consider $\beta=2.34$.  Note that $\beta$ represents the ratio of pipe radius to the correlation length-scale of the random walk in the absence of bias. Therefore, when modelling a random walk as a diffusion process, $\beta$ should be sufficiently large\cite{Bearon:2011}. However, we hypothesize that this restriction may be relaxed in the case of gyrotactic cells that are well-focussed by the flow along the axis of the tube and only suffer rare collisions with the wall.

\subsection{Calculation of mean swimming velocity and diffusion}
\citet{Hill:2002} demonstrate that the GTD equations (\ref{eq:f_eqn_with_flow}) and (\ref{eq:b_eqn_with_flow}) for  $f$ and $\mathbf{b}$, respectively, can in general be solved by expanding in spherical harmonics using a Galerkin method.  The method is summarized for the flow employed in this paper in appendix \ref{eq:App_Garlekin}.

To simplify the numerical solution of the swimming-advection-diffusion equation in pipe flow, we fit the rather complex
algebraic expressions in $\sigma$ obtained using the Galerkin method for the mean swimming direction and diffusion tensor  
with the simpler curves
\begin{eqnarray}
\label{eq:q_sigma_fit}
q^r(\sigma)&=&- \sigma P(\sigma; \mathbf{a}^r,\mathbf{b}^r),\quad q^z(\sigma)=-P(\sigma; \mathbf{a}^z, \mathbf{b}^z), \\
D^{rr}_G(\sigma)&=&P(\sigma; \mathbf{a}^{rr}_G, \mathbf{b}^{rr}_G),\quad
D^{rz}_G(\sigma)=- \sigma P(\sigma; \mathbf{a}^{rz}_G,  \mathbf{b}^{rz}_G), \quad
D^{zz}_G(\sigma)=P(\sigma; \mathbf{a}^{zz}_G, \mathbf{b}^{zz}_G),
\end{eqnarray}
where
\begin{eqnarray}
\label{eq:P_rat_func}
P(\sigma; \mathbf{a},\mathbf{b})=\frac{a_0 +a_2 \sigma ^2+a_4\sigma^4}{1+b_2 \sigma^2+b_4 \sigma^4}.
\end{eqnarray}
The choice of $ \mathbf{a}$ and $\mathbf{b}$ coefficients is described in table \ref{tab:func_fits} with reference to asymptotic results presented below. Please refer to appendix \ref{eq:App_functional fits} for the coefficients of fits to the full Galerkin solution.
We also consider results using the simpler estimate for the diffusion tensor, which we describe as the Fokker-Planck approximation (or FP), discussed in detail in appendix \ref{eq:App_FP_diff}.

\begin{table}
\caption{\label{tab:func_fits}In order to obtain the simplest functional fits whilst ensuring the asymptotic results are satisfied, the $ \mathbf{a}$  coefficients are as specified. The free parameters, $b_2, b_4$ are obtained  through least squared optimization of each fit of  velocity or diffusion component against $\sigma$. Because we are unable to obtain easily the coefficient of the $O(\sigma)$ correction to $D^{rz}_G$, in addition we allow $a^{rz}_{0,G}$ to vary. The fit coefficients for $\lambda=2.2$ are given explicitly in appendix \ref{eq:App_functional fits}. The subscript $G$ highlights that the results are for generalized Taylor dispersion (GTD), and ** indicates that the parameter is fitted.}
\begin{ruledtabular}
\begin{tabular}{l c c c}
 & $a_{0}$ &$a_{2}$ & $a_{4}$\\
\hline
$\mathbf{a}^r$		& $\frac{ J_1}{\lambda}$ 	&$\frac{2\lambda}{3} b^r_{4}$						&$0$\\
$\mathbf{a}^z$		&$K_1$			&$\frac{4 \lambda}{3} b^z_{4}$							&$0$\\
$ \mathbf{a}^{rr}_G$	&$\frac{ J_1}{\lambda^2}$	& $d_1b^{rr}_{4,G}$							&$0$\\
$ \mathbf{a}^{zz}_G$	&$\frac{ L_1}{\lambda}$	&$d_4b^{zz}_{4,G}+d_3b^{zz}_{2,G}$					&$d_3 b^{zz}_{4,G}$\\
$ \mathbf{a}^{rz}_G$	&**				&$d_2b^{rz}_{4,G}$							&$0$\\

\end{tabular}
\end{ruledtabular}
\end{table}

\begin{figure}[htb]
 \includegraphics[width=\textwidth]{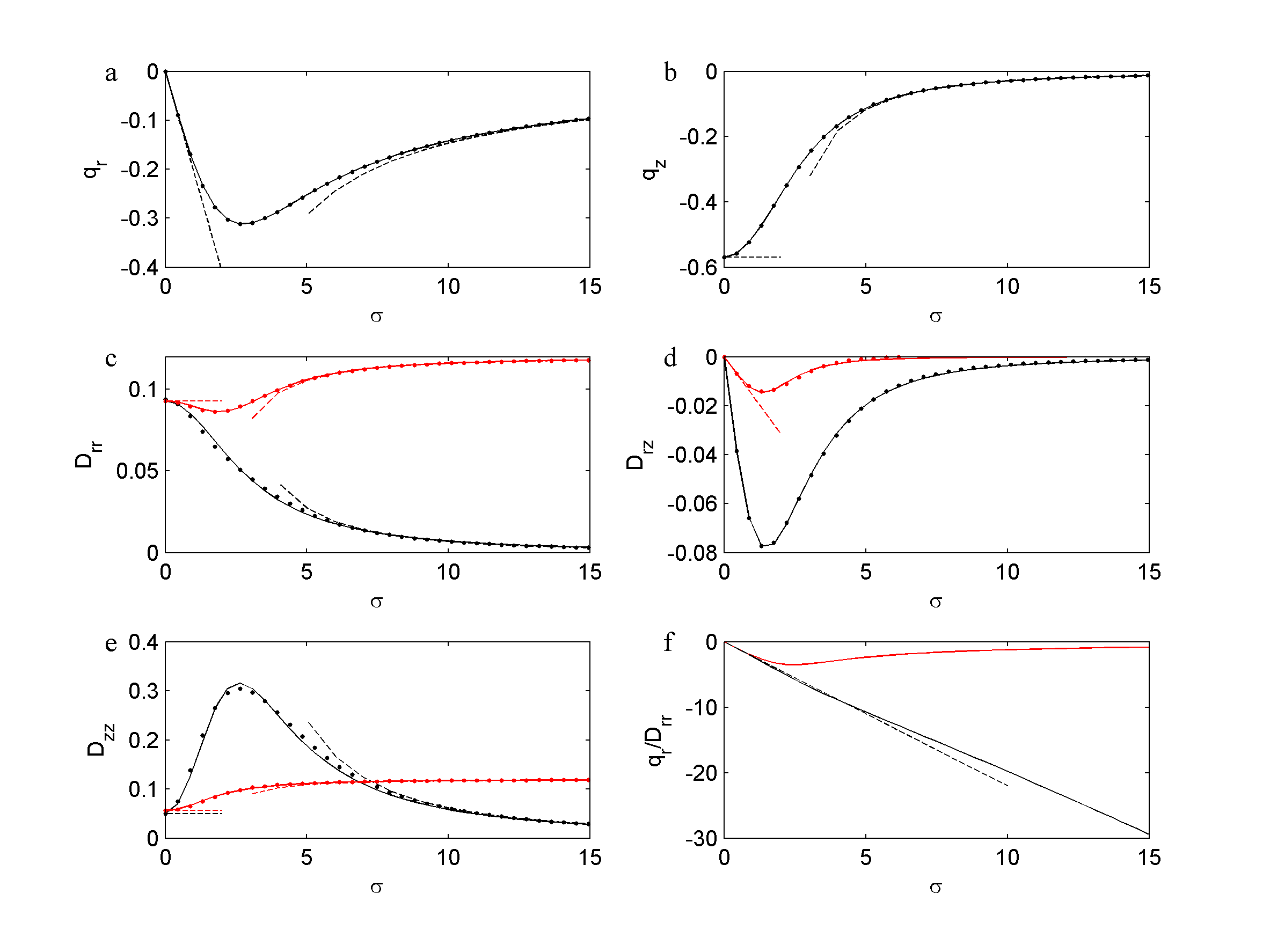}
  \caption{Mean swimming and diffusion coefficients as a function of shear, $\sigma$, for $\lambda=2.2$. Points are calculated using the Galerkin method, solid lines are functional fits described in the text, and dashed lines are asymptotic results. For diffusion calculations, black lines are for GTD, whereas red (grey) lines indicate the FP estimate.  }
  \label{fg:diffusion_fits}
  \end{figure}

In figure \ref{fg:diffusion_fits} we see that these simple functions are good approximations for the exact solutions for the mean swimming and diffusion, and it is evident how shear can significantly affect the mean swimming and diffusion. Furthermore, we note how that the calculation of diffusion via the GTD method is qualitatively different to that calculated via the simpler Fokker-Planck method (FP). In particular, we note that the components of diffusion approach zero in the limit of large shear using the GTD method, whereas they approach a finite non-zero limit via the FP method (see \citet{Hill:2002}).

To provide confidence in the results from the Galerkin method, we have obtained asymptotic expressions for $\sigma \ll 1$ and $\sigma \gg 1$, as described in appendices \ref{eq:App_small_sigma} and
\ref{eq:App_large_sigma}.

Specifically,  for $\sigma \ll 1$, the mean swimming direction with respect to 
coordinates ($\mathbf{e}_r,\mathbf{e}_\psi,\mathbf{e}_z$) correct to $O(\sigma)$ is given by
\begin{eqnarray}
\label{eq:mean_swim_small_sigma}
\mathbf{q}&=&-(\frac{\sigma}{\lambda} J_1,0, K_1)^T,
\end{eqnarray}
where the quantities $J_1$ and  $K_1$ are specified functions of  $\lambda$, coinciding with the results of \citet{Pedley:1990} using the FP model.  
However, calculation of the diffusion tensor from (\ref{eq:pos_def_diffusion}) reveals the new result that at leading order the diffusion tensor is diagonal with horizontal component $D^{rr}=\frac{J_1}{\lambda^2}$, and vertical component $D^{zz}=\frac{L_1}{\lambda}$, where $L_1$ is also a specified power series in $\lambda$ (appendix \ref{sec:small_sigma_D}). For $\lambda=2.2$ we have that $K_1=0.57, J_1=0.45$ and $L_1=0.11$, see appendix \ref{eq:App_small_sigma}. There is an $O(\sigma)$ correction to the off-diagonal term $D^{rz}$, but the second term in the definition of the diffusion tensor in (\ref{eq:pos_def_diffusion}) does not allow for a simple closed form expression for these components. (This is in contrast to expressions obtained by \citet{Pedley:1990} using the simpler orientation-only FP model with a diffusion estimate proportional to the variance of $\mathbf{p}$.)

For $\sigma \gg 1$, as in 
\citet{Bees:1998a},  
the mean swimming direction correct to $O(1/\sigma^2)$ is
\begin{eqnarray}
\mathbf{q}&=&-(\frac{2\lambda}{ 3\sigma},0,\frac{4\lambda}{ 3\sigma^2})^T.
\end{eqnarray}
Here, using GTD theory, we have the new result that the non-zero coefficients of the diffusion tensor are
\begin{eqnarray}
\mathbf{D}&=&
\left(
\begin{array}{ccc}
\frac{d_1}{\sigma^2}&0&-\frac{d_2}{ \sigma }\\
0&\frac{1}{6}-\frac{d_5}{\sigma^2}&0\\
-\frac{d_2}{ \sigma }&0&d_3+\frac{d_4}{\sigma^2}
\end{array}
\right),
\end{eqnarray}
where the quantities $d_1,d_2,d_3,d_4,d_5$ are polynomials in $\lambda$, given in appendix \ref{eq:App_large_sigma}. For $\lambda=2.2$,  we find that $d_1=0.68, d_2=0.0060,d_3=0.0020, d_4=5.9, d_5=1.3$.  

The asymptotic results are presented in figure \ref{fg:diffusion_fits}, indicating excellent agreement with results from the Galerkin method and demonstrating correspondence with the functional fits described above.

\section{Population-level numerical simulations}
\subsection{Numerical Methods}
 The governing swimming-advection-diffusion equation is solved using a
spatially adaptive finite element method as described in \citet{Bearon:2011}. The cell concentration, $n$, is approximated using standard
 Lagrangian quadratic finite elements and the time
 derivative is approximated using an implicit second-order, backward
 difference scheme. The subsequent discrete
 linear system is assembled using the C++ library
 \texttt{oomph-lib} \cite{Heil:2006} and solved by a direct solver,
 SuperLU \cite{Demmel:1999}.   In unsteady simulations, a fixed time-step of $dt=10^{-3}$ is used. The results were validated by repeating selected simulations with smaller error tolerances and timesteps.

\subsection{Steady gyrotactic focussing}
First, we seek an equilibrium solution $n(r)$ of equation (\ref{eq:nd_ad_diff_in_flow}) which represents gyrotactic focussing of cells towards the centre of the pipe.  Imposing zero flux on the pipe wall, at $r=1$, we have that
\begin{eqnarray}
 \beta q^r n-D^{rr}\frac{d n}{dr}=0,
\end{eqnarray}
which we can integrate to obtain
\begin{eqnarray}
\label{eq:equil_gyro}
n=n_0 \exp\left(\int  \frac{\beta q^r}{D^{rr}} dr\right),
\end{eqnarray}
where the radial components of the mean swimming direction and diffusion tensor, $q^r$ and $D^{rr}$, respectively, are functions of the local shear. In particular, if we take simple Poiseuille flow, $\chi(r)=1-2r^2$, we have that  $\sigma$,  the non-dimensional measure of the shear, is given by $\sigma=-\frac{Pe }{2 \beta^2} \chi'=\frac{2Pe }{ \beta^2} r$. For $\sigma\ll1$, at leading order we have that $q^r/D^{rr}=-\sigma \lambda=-\frac{2 Pe \lambda }{\beta^2} r$  from which we predict the Gaussian distribution
\begin{eqnarray}
\label{eq:Gauss_dist}
n=n_0 \exp\left(-\frac{Pe \lambda }{\beta} r^2\right).
\end{eqnarray}
As demonstrated in figure \ref{fg:diffusion_fits}(f), the leading order asymptotic solution $q^r/D^{rr}=-\sigma \lambda$ is an excellent approximation for $\sigma=O(1)$.  It is important to note that the GTD and FP methods yield a qualitative difference in the behaviour of $q^r/D^{rr}$.

\begin{figure}[htb]
 \includegraphics[width=0.8\textwidth]{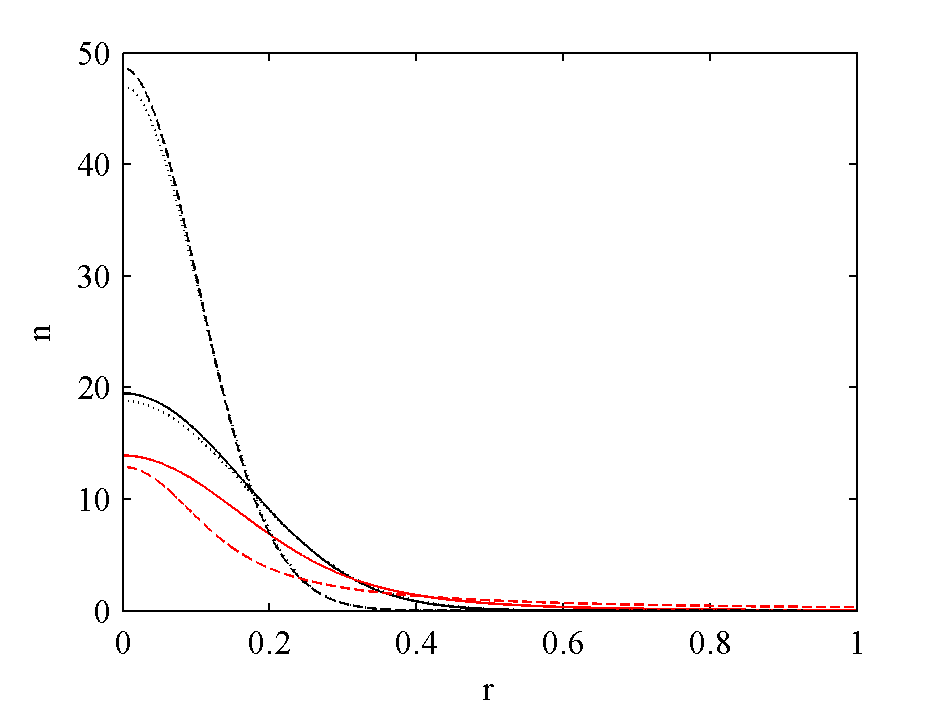}
  \caption{Equilibrium concentration (\ref{eq:equil_gyro})  for $Pe=20$ (solid line) and $Pe=50$ (dashed line) for swimming parameter $\beta=2.34$. Cell diffusion is calculated using GTD  (black) and FP (red/grey) approaches. The dotted lines are the associated Gaussian distributions  (\ref{eq:Gauss_dist}). The solutions are normalized so that there is unit total mass per unit length, $\int 2\pi r n(r) dr=1$. }
  \label{fg:equil_gyro_focus}
  \end{figure}

Example calculations of the equilibrium solution (Eq. \ref{eq:equil_gyro}) are shown in figure \ref{fg:equil_gyro_focus}. For the given values of $Pe$ and $\beta$, we see that cells undergo gyrotactic focussing, and that the distribution predicted by GTD theory can be well-approximated by the Gaussian distribution  (Eq. \ref{eq:Gauss_dist}) but shows a marked difference to that predicted by FP theory. Furthermore, whereas we see that GTD predicts an enhancement of gyrotactic focussing with increasing shear strength, the FP approximation predicts a reduction in gyrotactic focussing with increasing shear strength at sufficiently large shear.

\subsection{Vertical dispersion}
\begin{figure}[htb]
 \includegraphics[width=\textwidth]{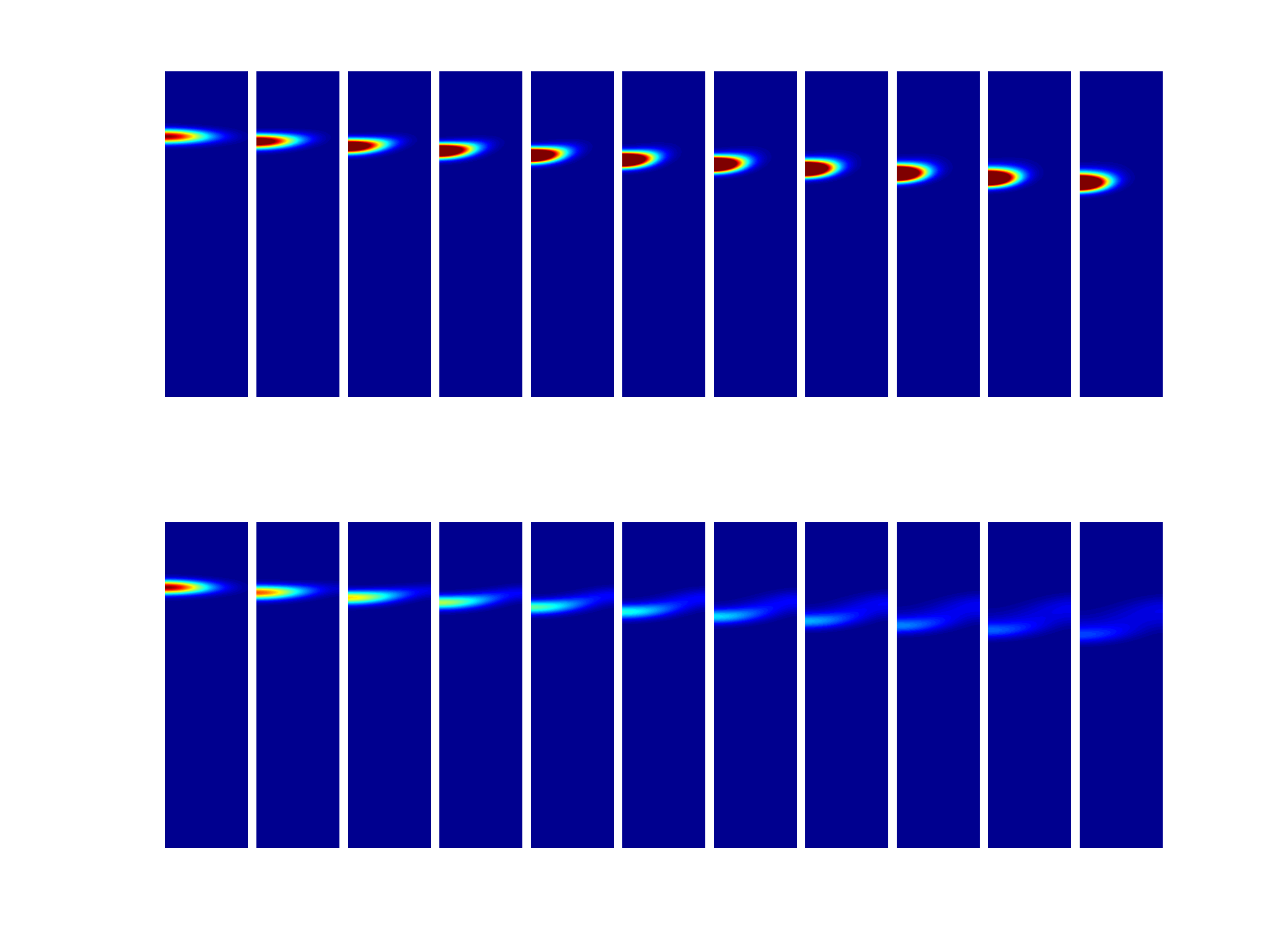}
  \caption{Concentration in region $z\in(0,600)$, $r\in (0,1)$ from $t=0$ to $t=1$ at intervals of $\delta t=0.1$ with $Pe=50$. Upper plots are for gyrotactic cells with $\lambda=2.2$, $\beta=10$. Lower plots are equivalent results for a passive solute. The colour scale is based on the initial concentration distribution, with red representing the maximal initial concentration at the centre of the blob of cells, and blue zero concentration}
  \label{fg:unsteady_comparison}
  \end{figure}

\begin{figure}[htb]
 \includegraphics[width=\textwidth]{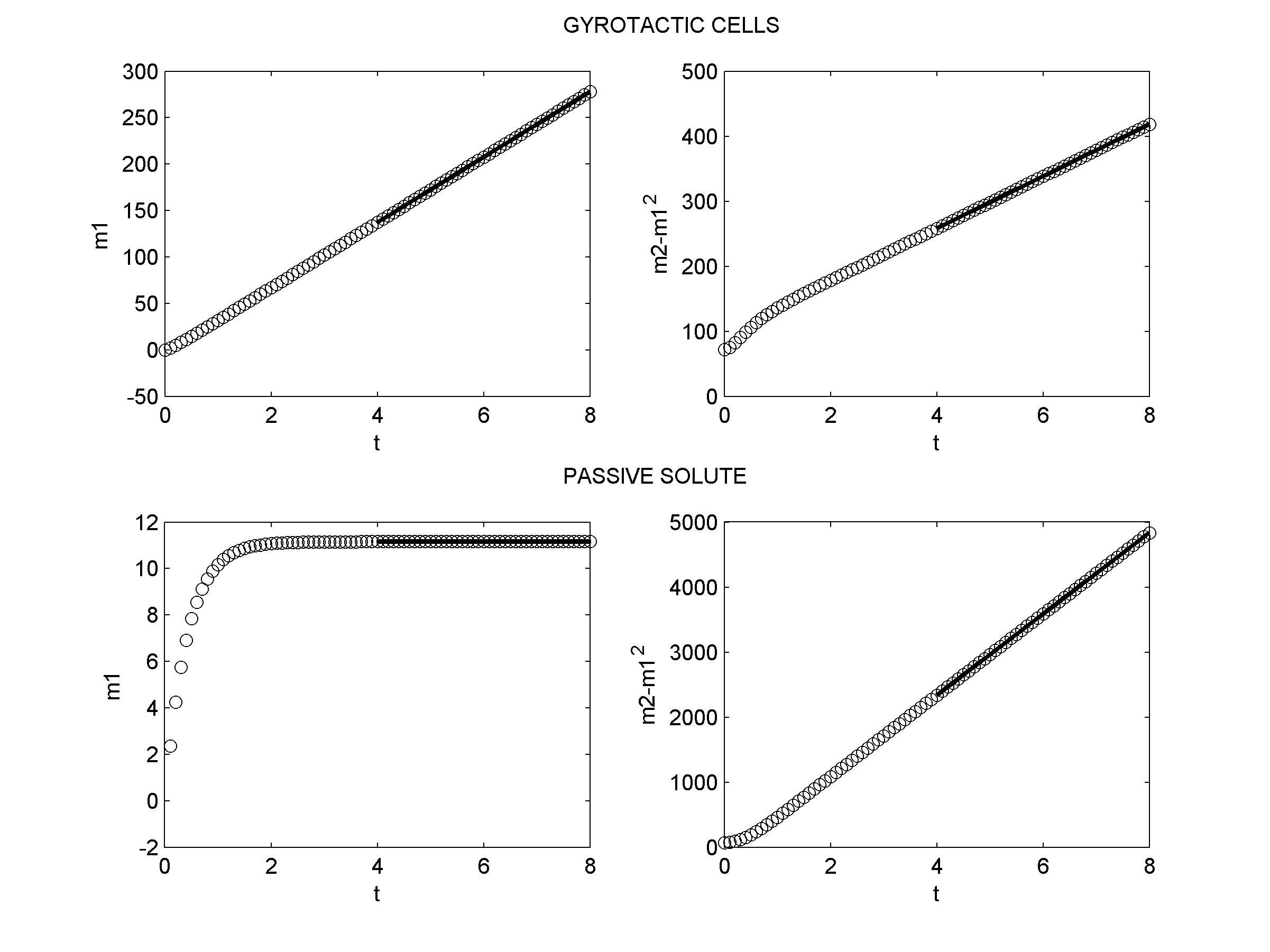}
  \caption{The mean and variance, $m_1$ and  $m_2-m_1^2$ of the distribution as a function of time, $t$ for $Pe=50$. Upper plots are for gyrotactic cells with $\lambda=2.2$, $\beta=10$. Lower plots are equivalent results for a passive solute. Open circles are results from numerical simulation, solid lines are linear regressions for $t\in[4,8]$. }
  \label{fg:drift_diffusion_transient}
  \end{figure}

\citet{Bees:2010} investigated how the average axial dispersion was modified for gyrotactic organisms compared with a passive solute. Specifically, using the method of moments and the FP approach they obtained long-time expressions for the vertical drift relative to the mean flow and the effective axial swimming diffusivity as a function of $Pe$ and a gyrotactic parameter.  Here, we perform a similar calculation, using simulations and the GTD calculations for the diffusion tensor.
We solve numerically the swimming-advection-diffusion equation (\ref{eq:nd_ad_diff_in_flow}) with initial condition
\begin{eqnarray}
n(r,z,0)=n_0  \exp\left(-\left(\frac{z-0.1L}{0.01 L}\right)^2-\left(\frac{r}{0.5}\right)^2 \right), 
\end{eqnarray}
representing a Gaussian blob of cells centred at $z=0.1L, r=0$. For the simulation domain we take $z\in (0,L)$, $r\in (0,1)$. 
Furthermore, we impose no-flux boundary conditions on the walls $r=1$, symmetry around the centreline, and periodic boundary conditions in the vertical direction, but take $L$ to be sufficiently large that boundary effects do not influence the vertical distribution. In the results presented in figures \ref{fg:unsteady_comparison} and \ref{fg:drift_diffusion_transient}, we take $Pe=50$, $L=1200$ and run the simulations for $t\in[0,8]$.  In figure \ref{fg:unsteady_comparison} we see example plots of the early concentration distribution as a function of time for both gyrotactic cells and a passive solute. For the passive solute, we take $\mathbf{D}=\frac{1}{6}\mathbf{I}$, and $\mathbf{q}=\mathbf{0}$ in equation  (\ref{eq:nd_ad_diff_in_flow}).  As shown in appendix \ref{eq:App_small_sigma}, this is equivalent to considering mean swimming and diffusion in the absence of gyrotactic bias, $\lambda=0$, and shear, $\sigma=0$. Whereas the gyrotactic cells are focussed towards the centre of the pipe, the passive solute diffuses radially.

Following \citet{Bees:2010}, we quantify dispersion in terms of cross-sectionally averaged axial moments of the concentration distribution.
To compute the moments of the distribution, we first translate to a reference frame moving with the mean flow, $\hat{z}=z-Pe~t$. The cross-sectional average, $m_p(t)$, of the $p$th axial moment, $c_p$, is (dropping hats for clarity)
\begin{eqnarray}
c_p(r,t) &=&\int z^pn (r, z,t) dz, \quad p=0,1,2, \\
m_p(t) &=&2 \int c_p (r,t) r dr, \quad p=0,1,2.
\end{eqnarray}
The mean and variance, $m_1$ and  $m_2-m_1^2$, of the distribution  are plotted in figure \ref{fg:drift_diffusion_transient}. The solution is normalised so that the total mass is unity, $m_0=1$.

From the calculations of the $m_1$ and $m_2$, we then define the axial drift and effective axial diffusion to be
\begin{eqnarray}
\Lambda_0 &=&\lim_{t\to\infty} \frac{d}{dt}m_1, \\
D_e &=&\lim_{t\to\infty} \frac{1}{2}\frac{d}{dt}(m_2-m_1^2).
\end{eqnarray}

As depicted in figure \ref{fg:drift_diffusion_transient}, for $Pe=50$, performing a linear regression over the interval $t\in[4,8]$ we obtain  $\Lambda_0 =35.2$ for the gyrotactic cells with parameters $\lambda=2.2$, $\beta=10$, compared to  the long-time limit of  $\Lambda_0=0$ for a passive scalar predicted from classic Taylor dispersion theory.  This occurs because gyrotactic cells are focussed towards the centre of the tube where the flow is fastest and, hence, they are transported more rapidly than the mean flow. Noting that $\mathbf{D}=\frac{1}{6}\mathbf{I}$ for the passive solute, for $Pe=50$ the classical Taylor dispersion result predicts that $D_e =1/6+6Pe^2/48=313$, which compares well with the numerical calculation of $D_e=312$.  For the gyrotactic cells with parameters $\lambda=2.2$, $\beta=10$, we see a much reduced axial dispersion, with an estimate of $D_e=20.0$.  As discussed by \citet{Bees:2010}, this reduction in axial dispersion can be explained due to gyrotactic focussing: by self-concentrating towards the axis of the tube, cells undergo a much reduced sampling of radial space and thus sidestep classical shear-induced Taylor dispersion. Furthermore, preliminary calculations based on equations 6.1 and 6.2 of \citet{Bees:2010} using the GTD values for the components of $\mathbf{q}$ and $\mathbf{D}$ give excellent agreement with these numerical computations. Specifically, the calculations yield $\Lambda_0 =35.2$ and $D_e=20.6$ for gyrotactic cells with parameters $\lambda=2.2$, $\beta=10$.

\section{Discussion}

Here, we have considered the spatial distribution of gyrotactic algae in axisymmetric pipe flow. We have computed a population-level swimming-advection-diffusion model where the mean swimming velocity and diffusion tensor are based on the local shear using the theory of generalized Taylor dispersion. We have shown how shear modifies the mean swimming velocity and diffusion tensor and, furthermore, demonstrated how the diffusion tensor differs qualitatively from previous simpler models, such as the ``Fokker-Plank" approach for which the diffusion tensor is estimated to be the product of the variance of the orientation distribution and a correlation timescale.  We have demonstrated that the shear-induced modification to mean swimming velocity and diffusion results in gyrotactic focussing and have quantified how the axial drift and diffusion of a population of cells is modified from that predicted for a passive scalar.

In this paper, we only considered unidirectional coupling between flow field and cell concentration. However, actively swimming cells, that are typically denser than the fluid, will modify the flow field. In a dilute suspension, where direct cell-cell hydrodynamic coupling can be neglected, negatively buoyant cells modify the flow field from Poiseuille flow \cite{Bees:2010}, which results in a change in local shear and thus a modification of the mean swimming velocity and diffusion tensor. Furthermore, direct hydrodynamic interactions between cells, and stresses induced by the swimming motions, may also alter the flow field \cite{Ishikawa:2009}.

Work in progress by the authors aims to incorporate the population-level model derived here from generalized Taylor dispersion in Bees and Croze's \cite{Bees:2010} modification of the classical Taylor-Aris theory in order to predict the axial drift and diffusion. Furthermore, both these predictions of long-time dispersion and the transient results presented in this paper will be compared with experimental observations of axial drift and diffusion in dyed suspensions of the alga {\it Dunaliella salina} in vertical tubes subject to imposed flow.  Finally, work is in progress by the authors to use direct numerical simulations to study the dispersion of active swimmers in laminar and turbulent flows, comparing statistical measures of dispersion from simulations with analytical predictions using the GTD expressions derived in this paper.

\begin{acknowledgments}
R.N.B. acknowledges assistance from A.L. Hazel to implement the C++ library \texttt{oomph-lib}.   M.A.B. and O.A.C. gratefully acknowledge support from  EPSRC (EP/D073398/1) and the Carnegie Trust.
\end{acknowledgments}

\appendix

\section{Galerkin method}
\label{eq:App_Garlekin}

To implement the Galerkin method, we follow the approach of \citet{Hill:2002} who considered the flow field $\mathbf{u}=G\zeta \mathbf{i}$.  Here, for the flow field $\mathbf{u}=G\xi \mathbf{k}$ (see Sec.~\ref{sec:vert}) we further extend the method to establish results for the full positive-definite diffusion tensor in (\ref{eq:pos_def_diffusion}).  We parameterize cell orientation in terms of spherical-polar co-ordinates ($\theta,\phi$)
\begin{eqnarray*}
\mathbf{p}=\sin\theta\cos\phi\mathbf{i}+\sin\theta\sin\phi\mathbf{j}+\cos\theta\mathbf{k}.
\end{eqnarray*}
Note that the direction $\theta=0$ corresponds to cells directed vertically upwards.

Equations (\ref{eq:f_eqn_with_flow}) and (\ref{eq:b_eqn_with_flow}) are solved by expanding $f$ and $b_j$, $j=1, 2, 3$, in spherical harmonics:
\begin{eqnarray}
\label{eq:gen_sigma_f_expan}
f&=&\sum_{n=0}^\infty\sum_{m=0}^nA_n^m\cos m\phi P_n^m(\cos\theta),\\
\label{eq:gen_sigma_b_expan}
b_j&=&\sum_{n=0}^\infty\sum_{m=0}^n(\beta_{nj}^m\cos m\phi +\gamma_{nj}^m\sin m\phi)P_n^m(\cos\theta).
\end{eqnarray}
Defining
\begin{eqnarray}
F_{nj}^m&\equiv& R_{nj}^m(\phi)P_n^m(\cos\theta),\mbox{~~~}j=0,1,2,3,\\
%B_{nj}^m&\equiv& R_{nj}^m(\phi)P_n^m(\cos\theta)
R_{nj}^m& =& \left\{ \begin{array}{ll} A_{n}^m \cos(m\phi), & j=0 \\
              \beta_{nj}^m \cos(m\phi) + \gamma_{nj}^m \sin(m\phi), \mbox{~~~} & j=1,2,3, \end{array} \right.
\end{eqnarray}
equations (\ref{eq:f_eqn_with_flow}) and (\ref{eq:b_eqn_with_flow}) then yield
\begin{eqnarray}
\sum_{n=0}^\infty\sum_{m=0}^n \left\{n(n+1) F_{nj}^m
+\lambda \sin^2\theta R_{nj}^m {P_n^m}'
+\sigma (\cos\phi \sin\theta R_{nj}^m {P_n^m}'
+\cot\theta\sin\phi {R_{nj}^m}' P_n^m) \right. \nonumber \\ \left.
-2\lambda \cos\theta F_{nj}^m \right\}
=\begin{cases}
0, & j=0, \\
\sum_{n=0}^\infty\sum_{m=0}^n (\sin\theta\cos\phi-(4\pi/3)A_1^1)F_{n0}^m, &j=1,\\
\sum_{n=0}^\infty\sum_{m=0}^n \sin\theta\sin\phi F_{n0}^m &j=2,\\
\sum_{n=0}^\infty\sum_{m=0}^n  \left\{ 2\sigma F_{n1}^m  +(\cos\theta-(4\pi/3)A_1^0)F_{n0}^m \right\}, &j=3,
 \end{cases}
\end{eqnarray}
where primes denote differentiation with respect to the dependent variable. Note that the normalization condition (Eq. \ref{eq:int_constraints}) requires that $A_0^0=1/4\pi, \beta_{0j}^0=0$, and from equation (\ref{eq:def_p}), we calculate the mean swimming direction to be $\mathbf{q}=(4\pi/3)(A_1^1,0,A_1^0)^T$. These equations can be simplified using identities for spherical harmonics so that inner products with other harmonics can be calculated. Finally, the resulting equations can be approximated by truncating the above series solutions to give a set of simultaneous equations that may be solved for the coefficients $A_n^m, \beta_{nj}^m$ and $\gamma_{nj}^m$.

The first term for the positive-definite diffusion tensor in equation (\ref{eq:pos_def_diffusion}), is given in part by equation $(52)$ of \citet{Hill:2002}, which only depends on the first few terms in the expansion (i.e.~$\beta_{1j}^m,\gamma_{1j}^m$, for $m=0,1$).  The second term cannot be written in such simple terms but can be approximated directly using all available coefficients.

\section{Small $\sigma$ asymptotics}
\label{eq:App_small_sigma}

The calculation of the mean swimming direction, $\mathbf{q}$, is the same for both the generalized Taylor dispersion theory and the Fokker-Planck approach\cite{Hill:2002}.
Hence, we follow \citet{Pedley:1990} to compute $f$ for small vorticity case (note that their small parameter $\epsilon$ is related to $\sigma$ via $\sigma=\epsilon \lambda$).

\subsection{Calculation of equilibrium distribution, $f$, and mean swimming, $\mathbf{q}$}

At leading order, $\sigma=0$, (\ref{eq:f_eqn_with_flow}) becomes
\begin{eqnarray*}
-\mathcal{L}_0f=
\frac{1}{\sin\theta}\frac{\partial}{\partial \theta}
\left(\sin\theta
\frac{\partial f}{\partial \theta}
\right)
+\frac{1}{\sin^2\theta}
\frac{\partial^2 f}{\partial \phi^2}
+\frac{\lambda}{\sin\theta}
\frac{\partial}{\partial \theta}
\left(
\sin^2\theta f
\right)=0.
\end{eqnarray*}
Looking for a solution independent of $\phi$, we obtain the von Mises distribution,
\begin{eqnarray}
f=f^{(0)}(\theta)=\mu e^{\lambda\cos\theta},
\end{eqnarray}
where (\ref{eq:fnorm}) yields the normalization constant $\mu=\lambda/4\pi\sinh\lambda$.

From equation (\ref{eq:def_p}), the mean swimming velocity is computed to be
\begin{eqnarray*}
\mathbf{q}^{(0)}=\int_0^{2\pi} \int_0^\pi \mathbf{p} f^{(0)}(\theta)  \sin\theta d\theta d\phi =(0,0,K_1),
\end{eqnarray*}
where $K_1=\coth\lambda -1/\lambda$.  For $\lambda=2.2$ we have $K_1=0.57$.

To find the $O(\sigma)$ correction, put
\begin{eqnarray*}
f=f^{(0)}(\theta)+\sigma f^{(1)},
\end{eqnarray*}
to obtain
\begin{eqnarray*}
-\mathcal{L}_0  f^{(1)}=
\frac{1}{\sin\theta}\frac{\partial}{\partial \theta}
\left(\sin\theta
\frac{\partial  f^{(1)}
}{\partial \theta}
\right)
+\frac{1}{\sin^2\theta}
\frac{\partial^2  f^{(1)}
}{\partial \phi^2}
+\frac{\lambda}{\sin\theta}
\frac{\partial}{\partial \theta}
\left(
\sin^2\theta  f^{(1)}
\right)=\lambda \cos\phi  \sin\theta  f^{(0)}.
\end{eqnarray*}
Looking for a solution of the form $f^{(1)}=\cos\phi F(\theta)$, we obtain the ODE
\begin{eqnarray}
\label{eq:ODE_F}
\frac{1}{\sin\theta}\frac{d}{d \theta}
\left(\sin\theta
\frac{d  F}{d \theta}
\right)
-\frac{F}{\sin^2\theta}
+\frac{\lambda}{\sin\theta}
\frac{d}{d \theta}
\left(
\sin^2\theta  F
\right)= \lambda \sin\theta  f^{(0)}.
\end{eqnarray}
Defining $x=\cos\theta$, and letting
\begin{eqnarray*}
F=-\mu  g_1(x),
\end{eqnarray*}
we obtain equation (3.4) of \citet{Pedley:1990}, which has a power series solution
\begin{eqnarray}
\label{eq:lambda_P11_exp}
g_1(x)=\sum_{n=1}^\infty \lambda^n A_n(x),\quad
A_n(x)=\sum_{r=1}^n a_{n,r} P^1_r(x),
\end{eqnarray}
where the $P^1_r$ are the associated Legendre functions and the coeffiicents $a_{n,r}$ satisfy
\begin{eqnarray*}
a_{n+1,r}=-a_{n,r+1}\frac{r+2}{(r+1)(2r+3)}+a_{n,r-1}\frac{(r-1)}{r(2r-1)}+\frac{e_{n+1,r}}{r(r+1)},
\end{eqnarray*}
where
\begin{eqnarray*}
e_{n+1,r}=\frac{2r+1}{n!2r(r+1)}\int_{-1}^1 (1-x^2)^{1/2} x^nP_r^1(x)dx.
\end{eqnarray*}

The first order correction to the mean swimming is then given by
\begin{eqnarray*}
\mathbf{q}^{(1)}=\int_0^{2\pi} \int_0^\pi \mathbf{p}  f^{(1)} \sin\theta d\theta d\phi = \int_0^{2\pi} \int_0^\pi \mathbf{p}\cos\phi  F(\theta)\sin\theta d\theta d\phi=(-\frac{J_1}{\lambda},0,0),
\end{eqnarray*}
where
\begin{eqnarray*}
J_1 =\frac{4\pi}{3}\lambda \mu \sum_{l=0}^\infty \lambda^{2l+1} a_{2l+1,1}.
\end{eqnarray*}
With $\lambda=2.2$ a calculation using Maple provides $J_1=0.45$.

Thus the  mean swimming correct to $O(\sigma)$ with respect to $\mathbf{i}, \mathbf{j}, \mathbf{k}$ unit vectors is given by $\mathbf{q}=(-\frac{\sigma}{\lambda} J_1,0, K_1)^T.$
Recalling the relationship between local and global co-ordinate vectors,
\begin{eqnarray}
\mathbf{i} =\mathbf{e}_r,\quad \mathbf{j}=-\mathbf{e}_\psi,\quad \mathbf{k}=-\mathbf{e}_z,
\end{eqnarray}
the mean swimming direction, with respect to $\mathbf{e}_r, \mathbf{e}_\psi,\mathbf{e}_z$  is (see Eq. \ref{eq:mean_swim_small_sigma})
\begin{eqnarray}
\mathbf{q}&=&-(\frac{\sigma}{\lambda} J_1,0, K_1)^T.
\end{eqnarray}

\subsection{Calculation of $\mathbf{b}$, and diffusion tensor, $\mathbf{D}$.}\label{sec:small_sigma_D}
At leading order, setting $\sigma=0$ in equation (\ref{eq:b_eqn_with_flow}) yields
\begin{eqnarray*}
-\mathcal{L}_0  \mathbf{b}=
\frac{1}{\sin\theta}\frac{\partial}{\partial \theta}
\left(\sin\theta
\frac{\partial \mathbf{b}}{\partial \theta}
\right)
+\frac{1}{\sin^2\theta}
\frac{\partial^2 \mathbf{b}}{\partial \phi^2}
+\frac{\lambda}{\sin\theta}
\frac{\partial}{\partial \theta}
\left(
\sin^2\theta \mathbf{b}
\right)=f^{(0)}(K_1\mathbf{k}-\mathbf{p}).
\end{eqnarray*}
By inspection, consider
\begin{eqnarray*}
b_\xi=B_H(\theta)\cos\phi\\
b_\eta=B_H(\theta)\sin\phi\\
b_\zeta=B_V(\theta).
\end{eqnarray*}
$B_H$ then satisfies the ODE
\begin{eqnarray}
\label{ref:hor_b}
\frac{1}{\sin\theta}\frac{d}{d \theta}
\left(\sin\theta
\frac{d B_H}{d \theta}
\right)
-\frac{B_H}{\sin^2\theta}
+\frac{\lambda}{\sin\theta}
\frac{d}{d \theta}
\left(
\sin^2\theta B_H
\right)
&=&-\sin\theta f^{(0)}.
\end{eqnarray}
On comparing equations (\ref{ref:hor_b}) and (\ref{eq:ODE_F}),  we can write
\begin{eqnarray*}
B_H=-\frac{1}{\lambda}F.
\end{eqnarray*}

From equation (\ref{eq:pos_def_diffusion}), the leading order expression for the non-dimensional horizontal component of diffusion ($D^{\xi\xi}=D^{\eta\eta}$) can thus be written as
\begin{eqnarray}
D^{\xi\xi}=\int_0^{2\pi} \int_0^\pi p_\xi \cos\phi B_H(\theta) \sin\theta  d\theta d\phi =-\frac{1}{\lambda}\int_0^{2\pi} \int_0^\pi p_\xi \cos\phi F(\theta) \sin\theta  d\theta d\phi =\frac{J_1}{\lambda^2}.
\end{eqnarray}

The function $B_V$ satisfies the ODE
\begin{eqnarray*}
\frac{1}{\sin\theta}\frac{d}{d \theta}
\left(\sin\theta
\frac{d B_V}{d \theta}
\right)
+\frac{\lambda}{\sin\theta}
\frac{d}{d \theta}
\left(
\sin^2\theta B_V
\right)
&=&(K_1-\cos\theta)f^{(0)}.
\end{eqnarray*}
To solve this, as for $F$, we define $x=\cos\theta$, let
\begin{eqnarray*}
B_V=\frac{\mu}{\lambda}  h_1(x),
\end{eqnarray*}
and seek power series solutions
\begin{eqnarray*}
h_1(x)=\sum_{n=1}^\infty \lambda^n B_n(x),\\
B_n(x)=\sum_{r=1}^n b_{n,r} P^0_r(x).
\end{eqnarray*}

By utilizing properties of Legendre polynomials we obtain the following recurrence relationship for the $b_{n,r}$:
\begin{eqnarray*}
b_{n+1,r}=-\frac{b_{n,r+1}}{2r+3}+\frac{b_{n,r-1}}{2r-1}+\frac{f_{n+1,r}}{r(r+1)},
\end{eqnarray*}
where
\begin{eqnarray*}
f_{n+1,r}=\frac{2r+1}{2n!}\int_{-1}^1 (x-K_1) x^nP_r^0(x)dx.
\end{eqnarray*}

The leading order expression for the non-dimensional vertical component of diffusion can thus be written as
\begin{eqnarray}
D^{\zeta\zeta}=\int_0^{2\pi} \int_0^\pi p_\zeta B_V(\theta) \sin\theta  d\theta d\phi = \frac{L_1}{\lambda},
\end{eqnarray}
where
\begin{eqnarray*}
L_1 =\frac{4\pi}{3}\mu \sum_{n=1}^\infty \lambda^{n} b_{n,1}.
\end{eqnarray*}
For $\lambda=2.2$ a computation employing Maple reveals that $L_1=0.11$.

The off-diagonal terms, $D^{\xi\zeta}$, etc., are all zero at leading order.

When comparing with a passive solute, we note that if we set $\sigma=0$, at leading order in  $\lambda$ we have that
\begin{eqnarray*}
\mu=1/4\pi, \quad J_1 =\frac{1}{3}\lambda^2 a_{1,1}, \quad  L_1 =\frac{1}{3} \lambda  b_{1,1}.
\end{eqnarray*}
Noting that $a_{1,1}=b_{1,1}=1/2$, it is clear that in the limit of $\lambda\to0$ the diffusion tensor tends to the isotropic tensor $\mathbf{I}/6$. This result can be obtained directly by considering equations (\ref{eq:f_eqn_with_flow}-\ref{eq:mathcalG}), which have solution $f=1/4\pi, \mathbf{b}=  \mathbf{p}/8\pi$.

\section{Large $\sigma$ asymptotics}
\label{eq:App_large_sigma}
For the calculation in the limit of large  $\sigma$, it is convenient to follow \citet{Manela:2003} and \citet{Brenner:1972} and define local co-ordinates so that the vorticity vector is in the direction of $\hat{ \mathbf{k}}$:
\begin{eqnarray}
\label{eq:local_co_ord2}
\hat{\mathbf{i}} =\mathbf{e}_r,\quad \hat{\mathbf{j}}=-\mathbf{e}_z,\quad\hat{ \mathbf{k}}=\mathbf{e}_\psi.
\end{eqnarray}

Defining the local position co-ordinate, $\mathbf{R}-\mathbf{R}_0=\xi\hat{\mathbf{i}} +\eta\hat{\mathbf{j}} +\zeta\hat{\mathbf{k}} $, the flow field can then be written locally as simple shear:
\begin{eqnarray}
\mathbf{u}(\mathbf{R})=\mathbf{u}(\mathbf{R}_0)+ G\xi\hat{\mathbf{j}},
\end{eqnarray}
where the shear strength $G$ is given as before by  $-\frac{U}{a}\chi'$.  For this flow field, the velocity gradient tensor has the simple form $G_{ij}=G \delta_{i1}\delta_{j2}$.

Writing the orientation vector as
\begin{eqnarray*}
\mathbf{p}=\sin\theta\cos\phi\hat{\mathbf{i}}+\sin\theta\sin\phi\hat{\mathbf{j}}+\cos\theta\hat{\mathbf{k}}
\end{eqnarray*}
we can write the governing equation (\ref{eq:f_eqn_with_flow}) as
\begin{eqnarray}
\label{eq:large_sigma_gov_eq}
\mathcal{L}f=
 \sigma\frac{\partial f}{\partial \phi}
-\mathcal{L}_s f=0
\end{eqnarray}
where the linear operator independent of $\sigma$ is given by
\begin{eqnarray*}
\mathcal{L}_s f=-\lambda\left(
\frac{1}{\sin\theta}
\frac{\partial}{\partial \theta}
\left(
\cos\theta\sin\theta\sin\phi f
\right)
+
\frac{\partial}{\partial \phi}
\left(\frac{\cos\phi}{\sin\theta} f
\right)\right)
+\nabla^2_{\mathbf{p}} f
\end{eqnarray*}
and where in sphericals the Laplacian is given by
\begin{eqnarray*}
\nabla^2_{\mathbf{p}}f=
\frac{1}{\sin\theta}\frac{\partial}{\partial \theta}
\left(\sin\theta
\frac{\partial f}{\partial \theta}
\right)
+\frac{1}{\sin^2\theta}
\frac{\partial^2 f}{\partial \phi^2}.
\end{eqnarray*}

For $\sigma\gg1$, we consider the following perturbation expansions for $f$ and $\mathbf{b}$:
\begin{eqnarray*}
f=\frac{1}{4\pi}\left(f^{(0)}+\frac{1}{\sigma}f^{(1)}+\left(\frac{1}{\sigma}\right)^2f^{(2)}+\dots\right)\\
\mathbf{b}=\frac{1}{4\pi}\left(\mathbf{b}^{(0)}+\frac{1}{\sigma}\mathbf{b}^{(1)}+\left(\frac{1}{\sigma}\right)^2\mathbf{b}^{(2)}+\dots\right).
\end{eqnarray*}

\subsection{Calculation of equilibrium distribution, $f$, and mean swimming, $\mathbf{q}$}

Substituting the expansion for  $f$ into equation (\ref{eq:large_sigma_gov_eq}) we obtain an explicit iterative scheme for computing the expansion:
\begin{eqnarray}
\label{eq:it_scheme}
\frac{\partial f^{(k+1)}}{\partial \phi}
&=&\mathcal{L}_s  f^{(k)}.
\end{eqnarray}

At leading order:
\begin{eqnarray*}
\frac{\partial f^{(0)}}{\partial \phi}
&=&0.
\end{eqnarray*}
Hence $ f^{(0)}= f^{(0)}(\theta)$ subject to $\int_0^\pi  f^{(0)}(\theta) \sin\theta d\theta=2$.

At $O(\frac{1}{\sigma})$:
\begin{eqnarray*}
\frac{\partial f^{(1)}}{\partial \phi}
&=&\mathcal{L}_s f^{(0)}(\theta).
\end{eqnarray*}
Because $ f^{(1)}$ must be periodic in $\phi$ with period $2\pi$, integrating this equation with respect to $\phi$ from $0$ to $2\pi$ gives:
 \begin{eqnarray*}
\frac{1}{\sin\theta}\frac{d}{d \theta}
\left(\sin\theta
\frac{df^{(0)}(\theta)}{d\theta}\right)
 =0.
\end{eqnarray*}
Excluding singular solutions, and given that $\int_0^\pi  f^{(0)}(\theta) \sin\theta d\theta=2$, we obtain the solution $ f^{(0)}=1$.

We can summarize the general iteration algorithm for the  terms $k\geq1$:
\begin{enumerate}
\item{
Integrate equation (\ref{eq:it_scheme})
\begin{eqnarray*}
 f^{(k+1)}&=&\int_0^\phi \mathcal{L}_s  f^{(k)}d\phi +F^{(k+1)}(\theta).
\end{eqnarray*}
}
\item{
Impose periodicity of $f^{(k+2)}$ \& integral constraint
\begin{eqnarray*}
\int_0^{2\pi} \mathcal{L}_s  f^{(k+1)}d\phi &=&0,\\
\int_0^{2\pi}  \int_0^{\pi} f^{(k+1)} d\theta d\phi &=&0,
\end{eqnarray*}
to determine non-singular solutions for $F^{(k+1)}(\theta)$.
}
\end{enumerate}

Specifically the first two terms in the expansions are given:
\begin{eqnarray*}
f^{(1)}&=&-2\lambda\cos\phi\sin\theta, \\
 f^{(2)}
&=&\frac{2}{3}\lambda^2(1-3\cos^2\theta)
+4\lambda\sin\theta\sin\phi
+\frac{3}{2}\lambda^2\cos 2\phi \sin^2\theta.
\end{eqnarray*}

From equation (\ref{eq:def_p}), we thus can compute  mean swimming at large $\sigma$ correct to $O(1/\sigma^2)$:
\begin{eqnarray}
q^\xi&=&\int_{\phi=0}^{2\pi}  \int_{\theta=0}^{\pi} f \sin^2\theta\cos\phi d\theta d\phi=-\frac{2\lambda}{ 3\sigma},\\
q^\eta&=&\int_{\phi=0}^{2\pi}  \int_{\theta=0}^{\pi} f \sin^2\theta\sin\phi d\theta d\phi=\frac{4\lambda}{ 3\sigma^2},\\
q^\zeta&=&\int_{\phi=0}^{2\pi}  \int_{\theta=0}^{\pi} f \cos\theta \sin\theta d\theta d\phi=0.
\end{eqnarray}
When converting back to the global co-ordinates we note that
\begin{eqnarray}
\hat{\mathbf{i}} =\mathbf{e}_r,\quad \hat{\mathbf{j}}=-\mathbf{e}_z,\quad\hat{ \mathbf{k}}=\mathbf{e}_\psi.
\end{eqnarray}
and so  with respect to $\mathbf{e}_r, \mathbf{e}_\psi,\mathbf{e}_z$ unit vectors  the mean swimming is given by
\begin{eqnarray}
\mathbf{q}&=&-(\frac{2\lambda}{ 3\sigma},0,\frac{4\lambda}{ 3\sigma^2})^T.
\end{eqnarray}

\subsection{Calculation of $\mathbf{b}$, and diffusion tensor, $\mathbf{D}$.}

We now apply similar techniques to calculate the vector field  $\mathbf{b}$. Substituting the expansion for  $\mathbf{b}$ into equation (\ref{eq:b_eqn_with_flow}) we obtain the following iterative scheme for computing the expansion:
\begin{eqnarray}
\label{eq:b_vector_it}
\frac{\partial \mathbf{b}^{(k+1)}}{\partial \phi}-2\mathbf{b}^{(k+1)}.\hat{\mathbf{G}}
&=&\left( 4\pi  (\mathbf{p}-\mathbf{q}) f \right)^{(k)}
+\mathcal{L}_s  \mathbf{b}^{(k)}.
\end{eqnarray}
For the simple shear flow with  $\hat{G_{ij}}= \delta_{i1}\delta_{j2}$, taking the dot product with $\hat{\mathbf{i}}$ yields:
\begin{eqnarray}
\label{eq:bx_it}
\frac{\partial b_\xi^{(k+1)}}{\partial \phi}&=&\left( 4\pi (\sin\theta\cos\phi-q^\xi) f \right)^{(k)}
+\mathcal{L}_s  b_\xi^{(k)}
\end{eqnarray}

The method follows as for $f$:
\begin{enumerate}
\item{
Integrate equation (\ref{eq:bx_it})
\begin{eqnarray*}
b_\xi^{(k+1)}&=&\int _0^\phi \left( 4\pi  (\sin\theta\cos\phi-q^\xi) f \right)^{(k)}
+\mathcal{L}_s  b_\xi^{(k)}d\phi + B_\xi^{(k+1)}(\theta).
\end{eqnarray*}
}
\item{
Impose periodicity of $b_\xi^{(k+2)}$ \& integral constraint
\begin{eqnarray*}
\int_0^{2\pi}\left(4\pi (\sin\theta\cos\phi-q^\xi) f \right)^{(k+1)} +\mathcal{L}_s  b_\xi^{(k+1)}d\phi &=&0,\\
\int_0^{2\pi}  \int_0^{\pi} b_\xi^{(k+1)} d\theta d\phi &=&0,
\end{eqnarray*}
to determine non-singular solutions for $B_\xi^{(k+1)}(\theta)$.
}
\end{enumerate}

We obtain the following expressions:
\begin{eqnarray*}
b_\xi^{(0)}&=&0,\\
 b_\xi^{(1)}&=&\lambda(1-3\cos^2\theta)/36+\sin\theta\sin\phi,\\
b_\xi^{(2)}&=&B_\xi^{(21)}(\theta)\cos\phi+B_\xi^{(22)}(\theta)\sin 2\phi.
\end{eqnarray*}

Taking the dot product of Eq. \ref{eq:b_vector_it} with  $\hat{\mathbf{j}}$ yields:
\begin{eqnarray}
\label{eq:by_it}
\frac{\partial b_\eta^{(k+1)}}{\partial \phi}&=&2b_\xi^{(k+1)}+\left( 4\pi (\sin\theta\sin\phi-q^\eta) f \right)^{(k)}
+\mathcal{L}_s  b_\eta^{(k)}\end{eqnarray}

\begin{enumerate}
\item{
Integrate Eq. \ref{eq:by_it}
\begin{eqnarray*}
b_\eta^{(k+1)}&=&\int _0^\phi 2b_\xi^{(k+1)}+\left( 4\pi (\sin\theta\sin\phi-q^\eta) f \right)^{(k)}
+\mathcal{L}_s  b_\eta^{(k)}d\phi + B_\eta^{(k+1)}(\theta).
\end{eqnarray*}
}
\item{
Impose periodicity of $b_\eta^{(k+2)}$ \& integral constraint
\begin{eqnarray*}
\int_0^{2\pi}2b_\xi^{(k+2)}+\left(4\pi (\sin\theta\sin\phi-q^\eta)  f\right)^{(k+1)}  + \mathcal{L}_s  b_\eta^{(k+1)}d\phi &=&0,\\
\int_0^{2\pi}  \int_0^{\pi} b_\eta^{(k+1)} d\theta d\phi &=&0,
\end{eqnarray*}
to determine non-singular solutions for $B_\eta^{(k+1)}(\theta)$. Note that calculation of $b_\eta^{(2)}$ requires calculation of $b_\xi^{(3)}$ which requires calculation of $f^{(3)}$. These calculations were performed using Maple, and files are available from the authors on request.
}
\end{enumerate}
We obtain the following expressions:
\begin{eqnarray*}
b_\eta^{(0)}&=&\lambda(1-3\cos^2\theta)/108\\
b_\eta^{(1)}&=&B_\eta^{(11)}(\theta)\cos\phi\\
b_\eta^{(2)}&=&B_\eta^{(10)}(\theta)+B_\eta^{(21)}(\theta)\sin\phi+B_\eta^{(22)}(\theta)\cos 2\phi
\end{eqnarray*}

Taking the dot product of equation (\ref{eq:b_vector_it}) with  $\hat{\mathbf{k}}$ yields:
\begin{eqnarray}
\label{eq:bz_it}
\frac{\partial b_\zeta^{(k+1)}}{\partial \phi}&=& 4\pi f ^{(k)} \cos\phi +\mathcal{L}_s  b_\zeta^{(k)},
\end{eqnarray}
which when combined with periodicity and the integral constrain yield the following expressions:
\begin{eqnarray*}
b_\zeta^{(0)}&=&\frac{1}{2}\cos\theta\\
 b_\zeta^{(1)}&=&-\frac{3}{4}\lambda\sin(2\theta)\cos\phi\\
b_\zeta^{(2)}&=&B_\zeta^{(10)}(\theta)+B_\zeta^{(21)}(\theta)\sin\phi+B_\zeta^{(22)}(\theta)\cos 2\phi
\end{eqnarray*}

From equation (\ref{eq:pos_def_diffusion}) we can now compute the diffusion tensor correct to $O(1/\sigma^2)$ :
\begin{eqnarray}
D^{\xi\xi}&=&\frac{1}{\sigma^2}(\frac{2}{3}+\frac{1}{270 }\lambda^2)\\
D^{\eta\eta}&=&\frac{\lambda^2}{2430}+\frac{1}{\sigma^2}(6-\frac{2\lambda^2}{243}-\frac{41\lambda^4}{25515})\\
D^{\zeta\zeta}&=&\frac{1}{6}-\frac{5}{18\sigma^2}\lambda^2\\
D^{\xi\eta}=D^{\eta\xi}&=&\frac{1}{810 \sigma}\lambda^2
\end{eqnarray}
all other entries are zero.
When converting back to the global co-ordinates we note that
\begin{eqnarray}
\hat{\mathbf{i}} =\mathbf{e}_r,\quad \hat{\mathbf{j}}=-\mathbf{e}_z,\quad\hat{ \mathbf{k}}=\mathbf{e}_\psi.
\end{eqnarray}
and so  with respect to $\mathbf{e}_r, \mathbf{e}_\psi,\mathbf{e}_z$ unit vectors  the  diffusion tensor is given by
\begin{eqnarray}
\mathbf{D}&=&
\left(
\begin{array}{ccc}
\frac{d_1}{\sigma^2}&0&-\frac{d_2}{ \sigma }\\
0&\frac{1}{6}-\frac{d_5}{\sigma^2}&0\\
-\frac{d_2}{ \sigma }&0&d_3+\frac{d_4}{\sigma^2}
\end{array}
\right),
\end{eqnarray}
where
\begin{eqnarray}
d_1&=&\frac{2}{3}+\frac{1}{270 }\lambda^2\\
d_2&=&\frac{\lambda^2}{810 }\\
d_3&=&\frac{\lambda^2}{2430}\\
d_4&=&6-\frac{2\lambda^2}{243}-\frac{41\lambda^4}{25515}\\
d_5&=&\frac{5}{18}\lambda^2.
\end{eqnarray}

\section{Fokker-Planck calculation of diffusion}
\label{eq:App_FP_diff}

For the Fokker-Planck approximation,\cite{Pedley:1990} the diffusion tensor non-dimensionalised on $V_s^2/d_r$ is given by
\begin{eqnarray}
\mathbf{D}_{F}=\tau d_r \int_\mathbf{p}(\mathbf{p}-\mathbf{q})^2  f(\mathbf{p} ) d\mathbf{p},
\end{eqnarray}
where $\tau$ is a directional correlation time estimated from experimental data. Although the quantity $\tau$ may vary with both $\lambda$ and the shear, $\sigma$,  for simplicity it is typically assumed to be independent of the shear. Asymptotic results for the diffusion tensor for weak shear\cite{Pedley:1990}  ($\sigma\ll1$) and strong shear \cite{Bees:1998a}  ($\sigma\gg1$) are available. With this choice of non-dimensionalisation and using the notation of this paper, the  $\sigma\ll1$ result correct to  $O(\sigma^2)$ is given by
\begin{eqnarray}
D^{rr}_{F}&=&\tau d_r \frac{K_1}{\lambda } ,\quad
D^{rz}_{F}=\tau d_r\frac{J_2-K_1J_1}{\lambda} \sigma,  \quad
D^{zz}_{F}=\tau d_r K_2 ,
\end{eqnarray}
where  $K_2$ and $J_2$ are specified functions of $\lambda$\cite{Pedley:1990}.The  $\sigma\gg1$ result correct to  $O(1/\sigma^3)$ is given by
\begin{eqnarray}
D^{rr}_{F}&=&\tau d_r \left(\frac{1}{3}-\frac{\lambda^2}{5 \sigma^2} \right),\quad
D^{rz}_{F}=0,  \quad
D^{zz}_{F}=\tau d_r \left(\frac{1}{3}-\frac{7\lambda^2}{45 \sigma^2} \right) .
\end{eqnarray}

To compare the Fokker-Planck approximation to the generalized Taylor  method we choose $\tau$ so that the two alternative calculations for the horizontal component of the diffusion agree when the shear is zero.  Specifically, when  $\sigma=0$ the generalized Taylor  method yields
$D^{rr}_G=\frac{J_1}{\lambda^2}$ and thus for the horizontal component of diffusion to agree we take $\tau  d_r= \frac{J_1}{\lambda K_1}$.   For the specific gyrotactic bias $\lambda=2.2$, this yields $\tau d_r =0.36$. Taking this value of $\tau$ also provides a value for the vertical component of diffusion, $D^{zz}_{F}=\tau d_r K_2  =0.056 $,  which is only a slight deviation from the generalized Taylor  method, $D^{zz}_G=  \frac{L_1}{\lambda}=  0.050$. Clearly, by this careful choice of $\tau$, the Fokker-Planck and generalized Taylor dispersion methods should agree when the shear is weak. However, as the shear increases we expect the two theories to give diverging predictions because, for example, in the FP approach $D^{rr}_F$ approaches $\frac{1}{3}\tau d_r$, whereas in the GTD approach, $D^{rr}_G$ tends to zero at large shear.

As for the generalized Taylor method, we fit simple functions to the curves of diffusion against $\sigma$ obtained with the Fokker-Planck method\cite{Pedley:1990,Bees:1998a}. The specific functions were given by
\begin{eqnarray}
\label{eq:q_sigma_fit_FP}
D^{rr}_{F}(\sigma)&=&P(\sigma; \mathbf{a}^{rr}_{F}, \mathbf{b}^{rr}_{F}),\quad
D^{rz}_{F}(\sigma)=- \sigma P(\sigma; \mathbf{a}^{rz}_{F},  \mathbf{b}^{rz}_{F}) \quad
D^{zz}_{F}(\sigma)=P(\sigma; \mathbf{a}^{zz}_{F}, \mathbf{b}^{zz}_{F}).
\end{eqnarray}
where the rational function $P(\sigma; \mathbf{a},\mathbf{b})$ is defined by equation \ref{eq:P_rat_func} and the choice of $ \mathbf{a}$ coefficients is described in table \ref{tab:func_fits_FP}.

\begin{table}
\caption{
\label{tab:func_fits_FP}
In order to obtain the simplest functional fits whilst ensuring the asymptotic results  are satisfied, the $ \mathbf{a}$  coefficients are as specified. }
\begin{ruledtabular}
\begin{tabular}{l||c|c|c}
 & $a_{0}$ &$a_{2}$ & $a_{4}$\\
\hline
$ \mathbf{a}^{rr}_{F}$	&$\frac{ J_1}{\lambda^2}$		& $-\frac{J_1\lambda}{5K_1}b^{rr}_{4,F}+\frac{J_1}{3K_1\lambda}b^{rr}_{2,F}$		 &$\frac{J_1}{3K_1\lambda}b^{rr}_{4,F}$\\
$ \mathbf{a}^{zz}_{F}$	&$\frac{ K_2J_1}{K_1\lambda}$	& $-\frac{7J_1\lambda}{45K_1}b^{zz}_{4,F}+\frac{J_1}{3K_1\lambda}b^{zz}_{2,F}$		 &$\frac{J_1}{3K_1\lambda}b^{rr}_{4,F}$\\
$ \mathbf{a}^{rz}_{F}$	&$\frac{(K_1J_1-J_2)J_1}{K_1\lambda^2}$		&$0$										&$0$
\end{tabular}
\end{ruledtabular}
\end{table}

\section{Coefficients of functional fits}
\label{eq:App_functional fits}

The fit coefficients for $\lambda=2.2$ for the mean swimming and diffusion are given by:

\begin{tabular}{l||c|c|c|c|c}
 & $a_{0}$ &$a_{2}$ & $a_{4}$&$b_{2}$ & $b_{4}$\\
\hline
$\mathbf{a}^r$		& $2.05 \times 10^{-1}$ 	&$1.86 \times 10^{-2}$	&$0$			&$1.74  \times10 ^{-1}$	&$1.27 \times 10^{-2}$\\
$\mathbf{a}^z$		&$5.7 \times 10^{-1}$		&$3.66 \times 10^{-2}$	&$0$			&$1.75 \times 10 ^{-1}$	&$1.25 \times 10^{-2}$\\
$ \mathbf{a}^{rr}_G$	&$9.30 \times 10^{-2}$	& $1.11 \times 10^{-4}$	&$0$			&$1.19\times 10 ^{-1}$	&$1.63 \times 10^{-4}$\\
$ \mathbf{a}^{zz}_G$	&$5.00 \times 10^{-2}$	&$1.11 \times 10^{-1}$	&$3.71 \times 10^{-5}$&$1.01 \times  10 ^{-1}$	&$ 1.86 \times 10^{-2}$\\
$ \mathbf{a}^{rz}_G$	&$9.17\times 10^{-2}$	&$1.56\times 10^{-4}$	&$0$			&$2.81 \times  10 ^{-1}$	&$2.62\times 10^{-2}	 $\\
$ \mathbf{a}^{rr}_{F}$	&$9.30 \times 10^{-2}$	& $5.73\times 10^{-4}$	&$1.85\times 10^{-3}$	&$4.96\times 10 ^{-2}$	&$1.54 \times 10^{-2}$\\
$ \mathbf{a}^{zz}_{F}$	&$5.60 \times 10^{-2}$	& $3.23 \times 10^{-2}$	&$1.70 \times 10^{-5}$&$2.70\times10 ^{-1}$		&$1.42 \times 10^{-4}$\\
$ \mathbf{a}^{rz}_{F}$	&$1.58 \times 10^{-2}$	&$0$				&$0$			&$9.61\times 10^{-2}$		&$7.88 \times 10^{-2}$
\end{tabular}

\nocite{*}

\bibliography{biblio23_02_12}
\end{document}